# CMOS-compatible athermal silicon microring resonators

Biswajeet Guha[1], Bernardo B. C. Kyotoku[1,2] & Michal Lipson[1]

[1]School of Electrical and Computer Engineering, Cornell University, Ithaca, New York 14853, USA.

[2]Departamento de Fisica, Universidade Federal de Pernambuco, Recife, Pernamubuco, Brazil.

**Silicon photonics promises to alleviate the bandwidth bottleneck of modern day computing systems[1,2,3]. But silicon photonic devices need to overcome the fundamental challenge of being highly sensitive to ambient temperature fluctuations due to the high thermo-optic (TO) coefficient of silicon (~$1.86 \times 10^{-4}$ K$^{-1}$)[4]. Most of the approaches proposed to date to overcome this problem either require significant power consumption or incorporate materials which are not CMOS-compatible. Here we demonstrate a new class of optical devices which are passively temperature compensated, based on tailoring the optical mode confinement in silicon waveguides. We demonstrate the operation of a silicon photonic resonator over very wide temperature range of greater than 80 degrees. The fundamental principle behind this work can be extended to photonic devices based on resonators such as modulators, routers, switches and filters.**

A passive CMOS-compatible thermal stabilization scheme for resonant photonic devices will go a long way towards enabling the implementation of ultralow power optical interconnects. This is because resonant devices such as microring resonators are ideally suited for dense integration of optical networks due to their compact size, high extinction ratio per unit length, low insertion loss and low power consumption[5,6,7]. But they are also highly sensitive to temperature (~ 0.11 nm/K) because of their narrow bandwidth. Typical temperature fluctuations in a commercial microprocessor can be 10s of degrees within a local hotspot[8], degrading considerably the performance of these resonant devices. There has been significant effort in stabilizing these devices by delocalizing the mode and overlaying a polymer coating with a negative TO coefficient[9,10,11,12], but polymers are currently not compatible with CMOS process. Temperature independent operation over a small range has been shown in multiple cavity coupled devices[13]. Another approach is to use local heating of silicon itself to dynamically compensate for any temperature fluctuations[14,15,16]. However, an active compensation scheme is both cumbersome (requiring thermo electric coolers and controllers) and power hungry.

Here we demonstrate the control of the thermal drift of photonic structures by tailoring the degree of optical confinement in silicon waveguides. The basic photonic structure we propose consists of a ring resonator overcoupled to a balanced Mach-Zehnder interferometer (MZI)[17, 18]. The schematic of the device is shown in Fig. 1a. The additional degree of freedom in the choice of waveguide widths[19], apart from just the lengths, enables one to set the thermal dependence of the MZI to counteract the thermal drift of the ring. The waveguide widths and lengths are chosen in the two arms of the MZI to give a balanced transmission ($\Delta(n \cdot L)_{MZI} = 0$) (see Fig. 1b) while having a strong negative temperature sensitivity overall ($\frac{\partial}{\partial T}\Delta(n \cdot L)_{MZI} < 0$). The ring has a large enough waveguide width to enable highly confined single mode operation, and consequently strong positive temperature sensitivity ($\frac{\partial}{\partial T}\Delta(n \cdot L)_{Ring} > 0$). The relative temperature sensitivities of the ring and the MZI, compared in Fig. 1c, are designed to cancel each other out.

The device is made inherently robust to temperature changes - its resonance oscillates about a central wavelength with temperature instead of drifting away, as is the case in standard resonators, due to the periodic interplay of the ring and MZI phase change with temperature. These oscillations arise due to the difference in the linear phase induced by the MZI with temperature and the nonlinear phase induced by the ring. This phase difference, captured in Eqn. (1), is converted into intensity modulation at the output of the structure.

$$\frac{\partial}{\partial T}\phi_{REMZI}(T) = \frac{\partial}{\partial T}\left\{phase\left(\frac{t - \alpha e^{j\beta L_{Ring}(T)}}{1 - \alpha t e^{j\beta L_{Ring}(T)}}\right)\right\} + \frac{\partial}{\partial T}(\beta L_{MZI}(T)) \qquad (1)$$

where $phase(X) = \arctan\left(\frac{\text{Im}(X)}{\text{Re}(X)}\right)$, X being a complex number, $L_{ring}$ and $L_{MZI}$ are the net optical path lengths for the ring and the MZI respectively, $t$ is the cross coupling coefficient of the ring to waveguide, and $(1-\alpha)$ is the roundtrip loss in the ring. The first term in Eqn. (1) is the change of phase induced by the ring with temperature, which redshifts the resonance with increase in temperature; while the second term refers to the change in optical path lengths of the two arms of the MZI with temperature designed to have a strong negative value, which compensates for the phase change of the ring. The difference between these phases can be seen in Fig. 2a where we plot, as an example, the phase change at a given wavelength of a 40 $\mu$m radius ring resonator as a function of temperature, and the phase added of the MZI at that wavelength. In all our simulations, we used thermo-optic coefficient of silicon as $1.86\times10^{-4}$ $K^{-1}$ and that of oxide as $1\times10^{-5}$ $K^{-1}$. It can be see that the nonlinearity of the ring phase gives rise to two distinct regions – one where the MZI added phase is smaller than the compensating phase required, and one where the MZI added phase is larger than the compensating phase required. The corresponding resonance lineshapes at each of these temperature ranges is shown in Fig. 2b. For a temperature change of $T_{per}$ the resonance lineshape exactly corresponds to the one at base temperature; hence $T_{per}$ is defined as one temperature cycle within which the resonance undergoes a complete oscillation as shown in Fig. 2b. It is this periodic mismatch between the phase added by the MZI and the phase compensation required that gives rise to oscillation in the ring resonance with temperature.

For a ring of given radius, only a particular choice of compensating MZI can result in perfect oscillations in the ring resonance with temperature. If the compensation is too large or too small, the ring resonance drifts away with increase of temperature while still exhibiting the periodic behaviour. This is in sharp contrast with normal ring resonator systems, where the resonance drifts monotonically with temperature. The locus of minima points of the spectra for such a system varies with temperature as

$$\Delta\lambda_{min} = (\frac{\lambda_0}{L_{Ring} + \chi L_{MZI}})\frac{\partial L_{Ring}}{\partial T}\left(T - \frac{2}{\beta\frac{\partial L_{Ring}}{\partial T}}\tan^{-1}\left\{\frac{1-t}{1+t}\tan(\frac{\gamma\beta\frac{\partial L_{Ring}}{\partial T}T}{2})\right\}\right) \quad (2)$$

where $\chi = \frac{1-t^2}{1+t^2+2t\cos(\beta L_{MZI})}$ and $\gamma = \frac{\partial L_{MZI}}{\partial T} / \frac{\partial L_{Ring}}{\partial T}$ is the compensation factor. Fig. 3a shows the behaviour of $\Delta\lambda_{min}$ for different cases of compensating factor ($\gamma$). Only when $\gamma = 1$, we get perfect oscillations in the resonance minima with temperature.

If $\gamma > 1$, the ring is overcompensated and the resonance slowly blueshifts with temperature; while if $\gamma < 1$, the ring is undercompensated and the resonance slowly redshifts. For reference $\Delta\lambda_{min}$ shift with temperature for a normal uncompensated ring resonator is also shown in Fig. 3a, which increases monotonically with temperature.

Smaller oscillations with temperature can be achieved by using structures where the phase compensation mismatch is smaller. This can be achieved using a structure with a large radius (see Fig. 3b). Since the resonances of a larger ring are closely spaced, the phase compensation mismatch width (shown in Fig. 2a) is smaller. Smaller oscillations can also be achieved using N smaller rings with radii $R \pm \delta r$ stabilized using one MZI (for WDM (wavelength division multiplexing) systems[20]). From Eqn. (2), it can be deduced that the resulting thermal oscillations ($\Delta\lambda_{min}$) will be $1/N$ times smaller than single ring case.

Devices were fabricated on a silicon-on-insulator (SOI) wafer with 240 nm Si thickness and 3 $\mu$m buried oxide thickness using electron-beam lithography. Fig. 4a shows the optical microscope image of a 40 $\mu$m radius ring resonator coupled to a balanced MZI, with SEM insets showing the corresponding waveguide widths. The wide and narrow waveguide widths were measured to be 420 nm and 190 nm respectively. The waveguides taper over a length of 10 $\mu$m at the width transition regions. The ring-to-waveguide coupling gap was 110 nm for this specific device. The lengths of the MZI arms are also shown in Fig. 4a. The measured quality factor of the ring was around 7000, a good value for switching and modulation applications for up to 40Gbps input data[21]. Quality factor is defined for $T=T_{per}$ (see Fig. 2c,).

The fabricated devices show temperature stability over a large temperature range of over 80K. Transmission spectra of this device at the bar port were measured at different temperatures. The transmission around 1565.5 nm for several different temperatures is shown in Fig 4b. For reference, the theoretical lineshapes at these temperatures are shown. The measured data agree very closely with the theoretical lineshapes. For a 40 $\mu$m radius silicon microring resonator, $T_{per}$ is around 22 $^o$C (since the free spectral range is ~2.2 nm in the C-band). In this particular case the oscillation in the wavelength at the transmission minima was less than 1 nm. We measured less than 3 dB worst case degradation in the transmission minima. Continuous operation over 80 degrees was demonstrated by passing a 1 Gbps, $2^7-1$ pseudo-random data at a bar port resonance of 1542.375 nm (Fig. 4c inset). The transmission wavelength was chosen slightly off-resonance at base temperature (22.5 C) since the spectral lineshape was not perfectly lorenztian. Eye patterns were obtained at different temperatures. Fig. 4c shows the eye patterns at different temperatures overlaid together, which clearly shows that the eye never closes at any temperature. Note that the rise time is not clearly visible due to resolution limitation. In fact the quality factor of these eye patterns[22] never goes below 10 with an error free operation (BER < $10^{-12}$)[23]. The eye opening decreases and increases with temperature as expected due to the oscillatory temperature dependence of the device.

In summary, we have demonstrated for the first time a temperature insensitive resonator-based device on silicon, with no extra power required for thermal stabilization. The device is shown to operate over a wide temperature and spectral range. The device is expected to have high performance as long as the dielectric refractive indices change linearly with temperature. It is also assumed that both the ring and the MZI are located in the same thermal hotspot which is typically $500 \times 500$ $\mu m^2$ in commercial microprocessors[8]. The approach presented here can be used in a ring resonator modulator system by surrounding the ring with diodes. A reduction in footprint of the device could be achieved by using narrower waveguides in MZI arm, or routing the arms in a coiled manner. The performance of the device could be further enhanced by using splitters/ couplers which are refractive index independent (like Y-splitters). This new generation of devices could lead to ultralow power on-chip optical interconnects capable of meeting the demands for the next generation of microprocessors.

**Acknowledgements** This work was performed in part at the Cornell Nano-scale Science & Technology Facility (a member of the National Nanofabrication Users Network) which is supported by National Science Foundation (Grant ECS-0335765), its users, Cornell University and Industrial Affiliates. We would also like to acknowledge the support of the National Science Foundation's CAREER Grant No. 0446571, as well as the Interconnect Focus Center Research Program at Cornell University, supported in part by Micro-Electronics Advanced Research Corporation (MARCO). The authors would also like to thank S. Manipatruni and L. Chen for helpful discussions.

**Author Contributions** B.G. designed, fabricated and tested the devices. B.B.C.K. helped in the testing. B.G., B.B.C.K. and M.L. discussed the results and their implications and contributed to writing this manuscript.

**Author Information** Correspondence and requests for materials should be addressed to M.L. (ml292@cornell.edu).

**Figure 1 Proposed device and its characteristics. a**, Schematic of the device showing the various waveguide lengths and widths. The MZI is highlighted in blue and the ring in red. **b**, Typical transmission spectrum for such a device with 40 $\mu$m ring radius and the MZI is balanced, i.e. the overall path lengths of the two arms are equal. **c**, Change in optical path length with temperature for the ring and MZI . The devices are designed to have opposite and equal phase shifts with increase in temperature.

**Figure 2 Oscillation of transmission resonance with temperature. a**, Phase change induced by in ring and by the MZI with temperature. The inherent nonlinearity in the ring phase gives rise to distinct overcompensated and uncompensated regions. This behaviour repeats itself after one temperature period $T_{per}$. **b,** Corresponding resonance lineshapes at different temperatures within one temperature period. The resonance displays periodic oscillations centred at $\lambda=\lambda_{res}$.

**Figure 3 Effect of compensation factor and ring radius**. **a**, Resonance minima shift for different cases of compensation for a ring resonator with 40 $\mu$m radius, showing the oscillatory behaviour with temperature. The monotonic drift of an uncompensated ring is also added for reference. **b**, Resonance minima shift with temperature for different ring resonator radii. The resonances oscillate less for larger rings, as compared to smaller rings.

**Figure 4 Proof of concept. a**, Optical microscope image of the device consisting of a 40 $\mu$m radius ring resonator coupled to a MZI whose lengths are shown. The coupling gap is 110 nm. SEM insets show the actual waveguide widths at various parts of the device. **b,** Bar port transmission spectrum of the device, centred around 1565.6 nm, at different temperatures. The dots represent actual measured data, and the straight lines represent theoretical lineshapes at those temperatures. **c**, Eye patterns of 1 Gbps input data at different temperatures overlaid. The probe wavelength was 1542.375 nm, which corresponds to a bar port resonance at base temperature (22.5 $^o$C). The eye-patterns show error free operation over around 80 degrees.

**Figure 1**

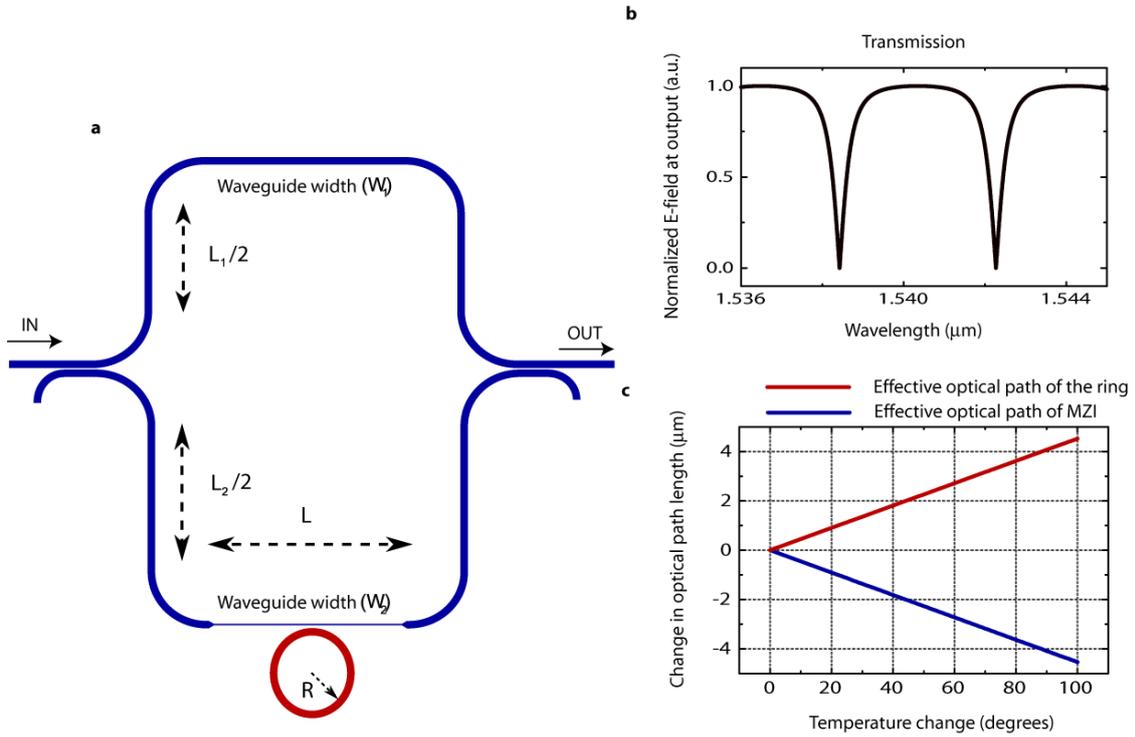

**Figure 2**

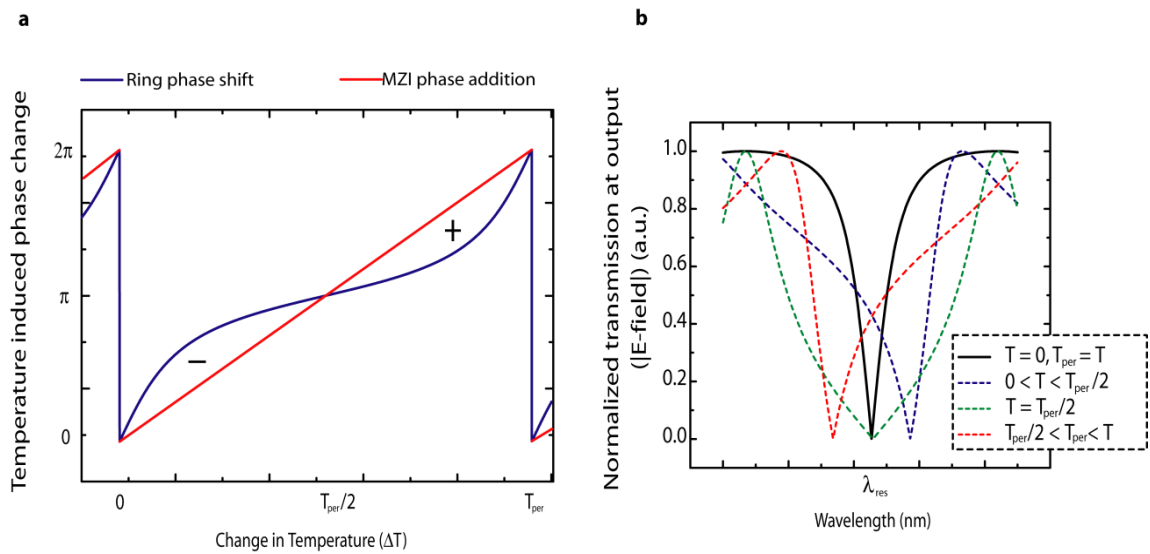

**Figure 3**

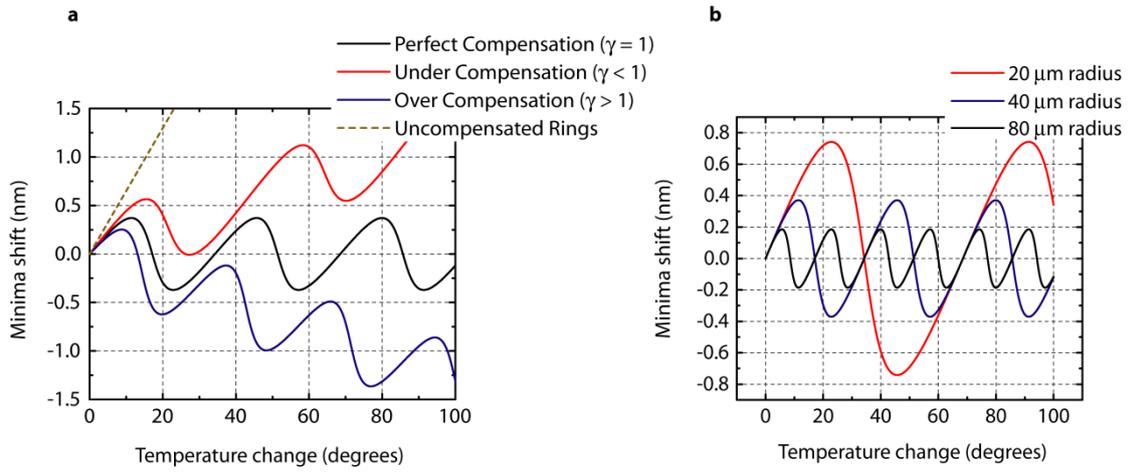

**Figure 4**

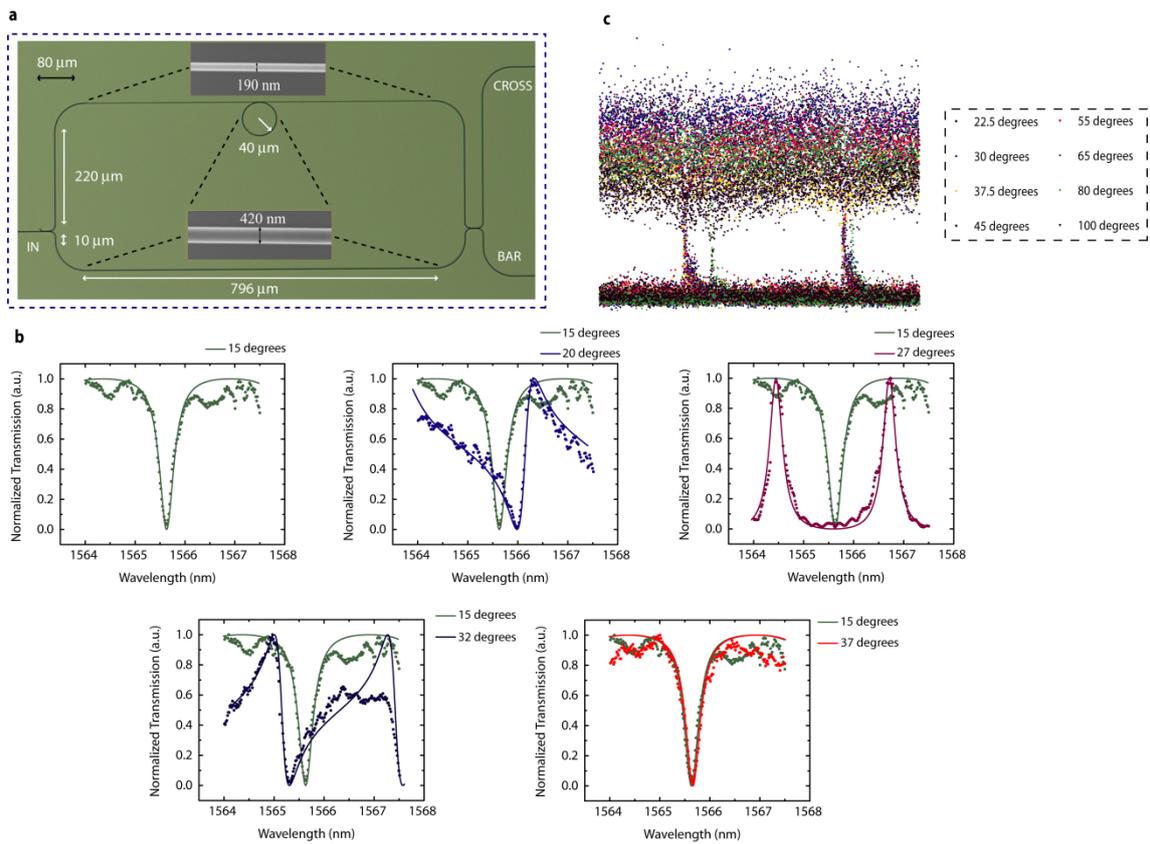